\documentstyle[12pt]{article}
\newcommand{\be}{\begin{equation}}
\newcommand{\ee}{\end{equation}}
\newcommand{\eee}{\end{eqnarray}}
\newcommand{\bee}{\begin{eqnarray}}

\newcommand{\ga}{\alpha}
\newcommand{\gb}{\beta}
\newcommand{\gga}{\gamma}
\newcommand{\gd}{\delta}
\newcommand{\gl}{\lambda}

\newcommand{\gep}{\epsilon}
\newcommand{\gvep}{\varepsilon}
\newcommand{\gs}{\sigma}
\newcommand{\go}{\omega}

\newcommand{\ty}{\tilde{y}}

\newcommand{\nn}{\nonumber}

\newcommand{\ua}{\underline{a}}
\newcommand{\ub}{\underline{b}}

\newcommand{\str}{{\rm str}}
\newcommand{\cS}{{\cal S}}

\begin{document}

\thispagestyle{empty}


\vspace*{2cm}

\begin{center}
{\large\bf  COORDINATE-FREE
ACTION FOR $AdS_3$ HIGHER-SPIN-MATTER SYSTEMS}\\
\vglue 1 true cm \vspace{2cm}
{\bf S.~F.~Prokushkin $^{1}$, A.~Yu.~Segal {{$^{2}$}}${\ }^{\dag}$ ,
and M.~A.~Vasiliev $^{3}$ }
\\
\vspace{1cm}

$^{1}$ Department of Physics, Stanford University,
       Stanford CA 94305-4060

\vspace{1cm}

$^{2,3}$ I.E.Tamm Department of Theoretical Physics, Lebedev Physics
Institute,\\
Leninsky prospect 53, 117924, Moscow

\vspace{1cm}

\vspace{1.5cm}
\end{center}

\begin{abstract}

A coordinate-free action principle for the
${\cal N}=2$ supersymmetric model of spin 0 and
spin 1/2 matter fields interacting via Chern-Simons
higher spin gauge fields in $AdS_3$ is formulated
in terms of star-product algebra of two oscillators.
Although describing proper relativistic dynamics
the action does not contain space-time coordinates.
It is given by a supertrace of a forth-order polynomial
in the $3d$ higher spin superalgebra. The theory admits a
free parameter characterizing the inequivalent vacua
associated with the parameter of mass $m$ of the matter fields.
For the case of the massless matter, the model has
a form of the noncommutative $2d$ Yang-Mills theory with some
infinite-dimensional gauge group. The limit $m\to\infty$
corresponds to the $SU(\infty)$ $2d$ Yang-Mills theory.

\end{abstract}

\vspace{1cm}

${\ }^{\dag}$
On leave of absence from Department of Physics, Tomsk State
University, \\ \indent {}
Lenin Ave., 36, Tomsk 634050, Russia

\newpage

\section{Introduction}

The pure gauge Chern-Simons higher spin
action in $AdS_3$ was constructed in \cite{Blen}.
Since the Chern-Simons
higher spin gauge fields do not propagate, this action
does not describe any nontrivial
local dynamics. So far, no attempt to analyze higher spin
gauge interactions of $3d$ matter fields at
the action level has been undertaken.
The nonlinear equations of motion have been constructed
however in all orders in interactions  for
$3d$ higher-spin-matter systems in \cite{Eq} for massless matter and
in \cite{PV} for an arbitrary mass of the matter fields. These
equations are equivalent to the Lagrangian ones at least at the
linearized level but
do not have manifestly Lagrangian form because of the presence of
an infinite
set of auxiliary variables. This makes it not straightforward
to derive the full nonlinear Lagrangian principle underlying the
higher spin dynamics
starting from this formulation of the higher spin equations.

The point is that, in this approach
(referred to as ``unfolded formulation" \cite{Unf}; also
see \cite{huge} and references therein)
 by virtue of introducing
an infinite set of auxiliary variables the equations of motion acquire
 ``unfolded'' form of certain zero-curvature equations and covariant
constancy conditions,
\be
\label{0cur}
   d\go =\go\wedge\go\,,\qquad
   dB^A =\go^i (t_i){}^A{}_B B^B\,,
\ee
supplemented with some gauge invariant constraints
\be
\label{con}
   \chi (B) =0
\ee
that do not contain spacetime derivatives.
Here $d=dx^{\ua} \frac{\partial}{\partial x^{\ua}} $
(spacetime base indices $\ua , \ub \ldots =0, \ldots, d-1$ are
underlined),
$\go (x)=dx^{\ua} \go_{\ua}^i (x)T_i $ is a gauge field
of some infinite-dimensional higher spin Lie superalgebra
$l$ ($T_i \in l$), and $B^A (x)$ is a set of 0-forms that take values
in the representation space of an appropriate infinite-dimensional
representation $(t_i ){}^B {}_A$ of $l$.
Note that in order
to reformulate a relativistic dynamics in such a form one
has to use an infinite-dimensional representation $t$ rich enough
to allow $B^A (x)$ to contain dynamical fields along with all their
on-mass-shell derivatives (for more detail see  \cite{huge}).
Such a form of the equations of motion is obviously
non-Lagrangian because
the number of equations exceeds greatly the number of variables.
It is likely that to construct the full non-linear action leading to
the equations (\ref{0cur}), (\ref{con})
 one has to add more auxiliary fields. This problem is not yet solved.

Since the equations (\ref{0cur}) have the form of zero-curvature
and covariant constancy conditions, they admit an explicit (local)
generic solution.  The dynamical content is therefore
hidden in the constraints (\ref{con}).
The role of the equations (\ref{con})
is that any their solution describes, at any point of space-time, the
complete set of all space-time derivatives of the dynamical fields,
which are allowed to be non-zero by the field equations
(for more detail see \cite{huge}). In this respect the
unfolded formulation formulation is analogous to the coordinate-free
approach of Penrose \cite{Pen,Penrose}.

The main observation of this paper is that
for the $3d$ higher spin model \cite{PV} it is possible to build an
action that leads to the constraints (\ref{con}) as its field
equations.  This action principle is formulated not in the usual
spacetime but in some auxiliary
noncommutative spinor space that describes the infinite-dimensional
representation space of $B^A$ as an appropriate functional space.
Remarkably, the constructed action has a form of that of the
2-dimensional noncommutative Yang-Mills theory in this auxiliary
space. Since the constraints (\ref{con}) encode the full information
about the dynamical system under investigation one can
adopt an extreme point of view that the proposed action gives just
the right variational principle for the model. As such,
the proposed model provides the first
example of the full nonlinear action principle for a nontrivial
$d=3$ model exhibiting higher spin gauge symmetries.

The paper is organized as follows. In sect.~\ref{eqs} we
collect following \cite{PV} some relevant facts about the
nonlinear system of equations for massive matter fields of spin 0
and 1/2 interacting via the Chern-Simons higher spin gauge potentials
in $AdS_3$. In sect.~\ref{act} we find the
action functional for these equations and show that in the massless
case it can be rewritten in a form analogous to the $2d$
noncommutative Yang-Mills action with some infinite-dimensional
gauge group.
In sect.~\ref{classes} we discuss a generalization of this
action to the case of an arbitrary mass.
In sect.~5 we focus on the special cases of the parameter of mass
at which the gauge algebra becomes finite-dimensional, as well as
on the quasiclassical limit.
In Conclusion we summarize the results and discuss some open problems.

\section{3D Higher-Spin-Matter Equations of Motion} \label{eqs}

The full nonlinear system of $3d$ higher-spin-matter equations,
which is a particular realization of the equations (\ref{0cur})
and (\ref{con}), is formulated \cite{PV} in terms of the generating
functions
$W(z,y;\psi_{1,2},k,\rho | x)$, $B(z,y;\psi_{1,2},k,\rho | x)$,
and $S_\ga(z,y;\psi_{1,2},k,\rho | x)$
that depend on the spacetime coordinates $x^{\ua}$
$(\ua=0,1,2)$, auxiliary commuting spinors $z_\ga$, $y_\ga$
$(\ga=1,2)$, $[y_\ga,y_\beta]=[z_\ga,z_\gb]=[z_\ga,y_\gb]=0$,
a pair of Clifford elements
$\{\psi_i,\psi_j\}=2\delta_{ij}$ $(i=1,2)$ that commute with
all other generating elements, and another pair of Clifford-type
elements $k$ and $\rho$ which have the following properties,
\be
\label{Klein}
   k^2=1\,,\: \rho^2 =1 \,,\: k\rho +\rho k =0\,,\:
   ky_\ga=-y_\ga k\,,\: kz_\ga=-z_\ga k\,, \:
   \rho y_\ga=y_\ga \rho \,,\: \rho z_\ga=z_\ga \rho\,.
\ee

The spacetime 1-form $W=dx^{\ua} W_{\ua}(z,y;\psi_{1,2},k,\rho | x)$,
\bee
\label{gexp}
   W_{\ua}(z,y;\psi_{1,2},k,\rho |x) &=& \sum_{A,B,C,D=0}^1
    \sum_{m,n=0}^\infty \!\frac 1{m!n!}W^{ABCD}_{\ua,\,
    {\ga_1}\ldots {\ga_m}{\gb_1}\ldots {\gb_n}}(x) \nn \\
    && {}\times
    k^A \rho^B \psi_1^C \psi_2^D  z^{\ga_1}\ldots z^{\ga_m}
    y^{\gb_1}\ldots y^{\gb_n} \,,
\eee
is the  generating function for higher spin gauge fields.
$B=B(z,y;\psi_{1,2},k,\rho | x)$
is the generating function for  the matter fields.
It admits an analogous expansion with the
component fields identified with the $3d$ matter fields and all their
on-mass-shell non-trivial derivatives. The field
$S_\ga(z,y;\psi_{1,2},k,\rho | x)$ is expressed in terms
of $B$ modulo gauge transformations by virtue of the field equations.
As explained in sect.~\ref{act} (see also \cite{huge}),
$S_\ga$ and $B$ can be interpreted as noncommutative connection and
field strength  in a two-dimensional noncommutative YM theory with
some infinite-dimensional gauge group.

The multispinorial coefficients in the expansions
like (\ref{gexp}) of the functions
$W_{\ua}$, $B$, and $S_\ga$ carry standard Grassmann parity:
they are even (odd) if the number of spinor indices is even (odd),
and are defined to commute with the generating elements
$z_\ga\,,\,y_\ga\,,\,k\,,\,\rho\,,\,\psi_{1,2}$.

The generating functions are treated as elements of the associative
algebra with the product law
\be
\label{prod}
   (f*g)(z,y)=\frac{1}{(2\pi)^2}\int d^2u d^2v \;\exp(iu_\ga v^\ga)
   f(z+u,y+u) g(z-v,y+v) \,
\ee
for any two
functions $f(z,y)$ and $g(z,y)$.
(The integration variables $u$ and $v$ are
required to satisfy the commutation relations similar to those of $y$
and $z$ in (\ref{Klein}). The spinorial indices are raised and
lowered by the $2\times 2$ symplectic form $\epsilon_{\ga\gb}$
according to
$u^\ga=\epsilon^{\ga\gb} u_\gb$,
$u_\ga=-\epsilon_{\ga\gb} u^\gb$, $\epsilon_{12}=\epsilon^{12} =1$.)
For generic elements
$f(z,y;\psi_{1,2},k,\rho) = F(z,y)\Phi(\psi_{1,2},k,\rho)$ and
$g(z,y;\psi_{1,2},k,\rho) = G(z,y)\Psi(\psi_{1,2},k,\rho)$
the order of the product factors in the integrand of (\ref{prod})
is essential as the elements $\psi_1\,,\,\psi_2\,,\,k\,,\,\rho$
do not commute. Due to the sign changes in (\ref{Klein}) we have
\be
    \Phi(\psi_{1,2},k,\rho) G(z,y) = \tilde{G}(z,y)
    \Phi(\psi_{1,2},k,\rho)
\ee
with some $\tilde{G}$. The full product is then defined as
\be \label{prod1}
   (f*g)(z,y;\psi_{1,2},k,\rho) = (F*\tilde{G}) (z,y) \;
   \Phi(\psi_{1,2},k,\rho ) \Psi(\psi_{1,2},k,\rho )\,.
\ee

The product law (\ref{prod}) yields a particular realization of
the Weyl algebra,
\be
\label{zz}
[y_\ga,y_\gb]_*=-[z_\ga,z_\gb]_*=2i\epsilon_{\ga\gb},\;\;
[y_\ga,z_\gb]_*=0
\ee
($[a,b]_* = a * b - b * a$, $\{a,b\}_* = a * b + b * a$).

The generating functions therefore can be treated as elements of
the associative algebra with the product law (\ref{prod}),
(\ref{prod1}).

The full system of equations of motion for the system of $3d$
massive spin-0 and spin-1/2 matter interacting via higher spin
gauge fields has the form \cite{PV}
\bee
\label{WW}
    {}&&  dW=W*\wedge W \,, \\
\label{WB}
    {}&&  dB=W*B-B*W   \,,   \\
\label{WS}
    {}&&  dS_\ga=W*S_\ga-S_\ga*W   \,, \\
\label{SS}
    {}&&  S_\ga * S^\ga=-2i(1+B)   \,, \\
\label{SB}
    {}&&  S_\ga*B+B*S_\ga =0\,,
\eee
where  $W$ and $B$ are independent of $\rho$ while
$S_\ga$ is linear in $\rho$,
\be
\label{tru}
    W(z,y;\psi_{1,2},k,-\rho | x) = W(z,y;\psi_{1,2},k,\rho | x)\,,
    \quad
    B(z,y;\psi_{1,2},k,-\rho | x) = B(z,y;\psi_{1,2},k,\rho | x)\,,
\ee
\be
\label{rho}
    S_\ga(z,y;\psi_{1,2},k,-\rho | x)=
      -S_\ga(z,y;\psi_{1,2},k,\rho | x)\,.
\ee
Eqs.(\ref{WW})-(\ref{SB})
are general coordinate invariant (because of using the exterior
algebra formalism) and invariant under the infinitesimal higher spin
gauge transformations
\be
\label{delta W}
    \delta W = d\gvep - W * \gvep + \gvep * W  \,,\quad
    \delta B = \gvep * B - B * \gvep  \,,\quad
    \delta S_\ga = \gvep * S_\ga - S_\ga * \gvep \,,
\ee
where $\gvep=\gvep(z,y;\psi_{1,2},k | x)$ is an arbitrary gauge
parameter.

A particular vacuum solution of the system (\ref{WW})-(\ref{SB})
can be chosen in the form \cite{PV}
\bee
\label{vacua}
  B_0=\nu\, K  \,, \qquad S_{0\ga} = s_{0\ga} \,, \qquad
  W_0=\frac1{8i}\,(\go^{\ga\gb} + \gl h^{\ga\gb} \psi_1)
    \{\ty_\ga, \ty_\gb\}_*\,,
\eee
where $\nu$ is a constant parameter,
\be
\label{K}
K=k e^{i(zy)},\qquad (zy)=z_\ga y^\ga\,,
\ee
\be
\label{tilde z}
   s_{0\ga} (z,y) = \rho \left(z_\ga+\nu (z_\ga + y_\ga)
    \int_0^1 dtt e^{it(zy)} k \right) \,,
\ee
\be
\label{tilde y}
   \ty_\ga (z,y) = y_\ga+\nu (z_\ga + y_\ga)\int_0^1dt(t-1)
   e^{it(zy)}k \,,
\ee
 and $\go^{\ga\gb}$ and $h^{\ga\gb}$ are some
 $\ga \gb$-symmetric 1-forms  on $AdS_3$
satisfying the equations

\bee
\label{adseq}
   d\go_{\ga\gb}=\go_{\ga\gga}\wedge\go_\gb{}^\gga
   + \gl^2h_{\ga\gga}\wedge h_\gb{}^\gga \,, \qquad
    dh_{\ga\gb}=\go_{\ga\gga}\wedge h_\gb{}^\gga
    + \go_{\gb\gga}\wedge h_\ga{}^\gga \,.
\eee
Provided that $h_{\ua}{}^{\ga\gb}$ is a non-degenerate $3\times 3$
matrix, $\go^{\ga\gb}$ and $h^{\ga\gb}$ are identified with the
background $AdS_3$ Lorentz connection and frame field, respectively.
The parameter
$\gl$ is identified with the inverse radius of $AdS_3$.

As shown in \cite{PV}, the elements $s_{0\ga}(z,y)$ and
$\ty_\ga (z,y)$ commute to each other,
\be
\label{zy}
  [s_{0\ga}, \ty_\gb]_* = 0 \,,
\ee
and obey the commutation relations of the ``deformed oscillator''
algebra \cite{Wig,Quant},
\be
\label{y com}
   [\ty_\ga,\ty_\gb]_* = 2i\gep_{\ga\gb}(1+\nu k)\,,\quad
        \ty_\ga k=-k \ty_\ga \,;
\ee
\be
\label{z com}
   [s_{0\ga}, s_{0\gb}]_* = -2i\gep_{\ga\gb}(1+\nu K)\,,\quad
    s_{0\ga} * K = -K * s_{0\ga} \,;
\ee
\be
\label{K com}
 k^2=  K * K = 1.
\ee
These properties guarantee that the substitution of the
ansatz (\ref{vacua}) into (\ref{WW}) leads to the
$AdS_3$ equations (\ref{adseq}).

Fixation of the vacuum solution (\ref{vacua})-(\ref{adseq})
breaks the local symmetry (\ref{delta W}) down to some global
symmetry, the symmetry of the vacuum. It is generated by
a parameter $\gvep_{gl}(x)$ obeying the conditions
\be
\label{e gl1}
    d\gvep_{gl}=[W_0\,,\,\gvep_{gl}]_*\,,
\ee
\be
\label{e gl2}
    [\gvep_{gl}\,,\,S_{0\ga}]_*=0  \,,
\ee
which follow from the requirements that $\gd W_0=0$ and $\gd S_{0\ga}=0$
($\gd B_0=0$ holds trivially), i.e. $\gvep_{gl}$ belongs to the stability
subalgebra of the vacuum solution.
The condition (\ref{e gl2}) implies that $\gvep_{gl}$ belongs to
the centralizer subalgebra for the element $s_{0\ga}$ (\ref{tilde z}),
an infinite dimensional algebra $l^g$ spanned by the elements
$\ty_\ga$, $k$, $\psi_1$ \cite{PV}. The equation (\ref{e gl1})
fixes a dependence
of $\gvep_{gl}$ on the space-time coordinates $x^{\ua}$ in terms of
the initial data $\gvep_{gl}(x_0)=\gvep_{gl}^0$ at any space-time
point $x_0$ (in some neighborhood of $x_0$).
$l^g$ contains the $3d$ $N2$ SUSY superalgebra $osp(2,2)\oplus osp(2,2)$
as a finite dimensional subalgebra
spanned by generators $\Pi_{\pm}T^A$, where
\be
\label{Pi}
    \Pi_{\pm}=\frac{1\pm\psi_1}2
\ee
and $T^A=\{T_{\ga\gb}\,,\,Q_\ga^{(1)}\,,\,Q_\ga^{(2)}\,,\,J \}$ with
\be
\label{N2}
  T_{\ga\gb} = \frac1{4i}\{\ty_\ga,\ty_\gb \}_*\,,\quad
  Q_\ga^{(1)} = \ty_\ga     \,,\quad
  Q_\ga^{(2)} = \ty_\ga k   \,,\quad
  J = k+\nu \,.
\ee
The fact that these generators form $osp(2,2)$
is a simple consequence \cite{BDV} of the properties of the deformed oscillator
algebra (\ref{y com}). Thus, the system (\ref{WW})-(\ref{SB})
possesses $N=2$ global SUSY for arbitrary $\nu$.
Perturbations near the vacuum (\ref{vacua}) describe
a massive scalar $AdS_3$ hypermultiplet interacting
via Chern-Simons type higher spin gauge fields \cite{PV}.
The values of mass are related to the parameter $\nu$ as follows:
$m^2_\pm =\gl^2 \nu(\nu\mp 2)/2$ for bosons and
$m^2_\pm =\gl^2 \nu^2/2$ for fermions.

Since Eq.~(\ref{WW}) and Eqs.~(\ref{WB}), (\ref{WS})
have, respectively, a form of zero-curvature equation
and covariant constancy conditions
they admit a (local) generic solution of the form
\be
\label{resolution}
   W = dg(x) * g^{-1}(x), \quad B(x) = g(x) * b * g^{-1}(x),
   \quad S_\ga(x) = g(x) * s_\ga * g^{-1}(x) \,,
\ee
where $g(z,y;\psi_{1,2}, k | x)$
is an arbitrary invertible element, $g*g^{-1} = g^{-1} *g =1$,
while $b(z,y;\psi_{1,2},k)$ and $s_\ga(z,y;\psi_{1,2},k,\rho)$
are arbitrary $x$-independent elements.

Plugging these expressions into (\ref{SS}) and (\ref{SB}) one
obtains, as a consequence of the gauge invariance,
\bee
\label{SS1}
    {}&&  s_\ga * s^\ga = -2i(1+b)   \,, \\
\label{SB1}
    {}&&  s_\ga * b + b * s_\ga  =0\,.
\eee
Since these equations contain no dependence on the spacetime
coordinates the dynamical system under consideration turns out
to be entirely  reformulated  in terms of the
auxiliary space of spinor variables, noncommutative spinor space.
Recall that the role of the equations (\ref{SS1}) and   (\ref{SB1})
is that any their solution describes, at any point of space-time,
the complete set of all space-time derivatives of the dynamical
fields, which are allowed to be non-zero by the field equations
(for more detail see \cite{huge}).

\section{Coordinate-Free Action for the Massless Case
and Non-Commutative Yang-Mills Theory} \label{act}

The main observation of this paper is that the system of equations
(\ref{SS1}), (\ref{SB1}) admits a Lagrangian formulation with
the action of the form
\be
\label{action1}
   \cS = \str \; L \,,
\ee
where $L$ is some star-product Lagrangian
function of $b$ and $s_\ga$ specified below
and  $\str$ is the
supertrace operation  \cite{Fortsch}
which for the particular star-product (\ref{prod}) has the form
\cite{huge}
\be
\label{str}
   \str \; f(z,y) = \frac{1}{(2\pi)^2}\int d^2z\, d^2y \;
      \exp(-iz_\ga y^\ga) f(z,y) \,.
\ee
Since the algebra of functions $f(z,y;k,\rho,\psi_{1,2})$
is in fact  a tensor product
of the Weyl algebra with some matrix algebra, this definition
admits a unique extension to the elements dependent on
$k,\rho$ and $\psi_{1,2}$ with ${\rm str}\; f=0$ for any $f$
proportional to the first power of any of the variables
$k,\rho$ or $\psi_{1,2}$.

The defining property of the supertrace (\ref{str}) that fixes it
uniquely \cite{Quant} for the class of polynomial functions
$f(z,y;k,\rho,\psi_{1,2})$ is
\be
\label{prstr}
    \str \,(f*g)=(-1)^{\pi (f)}\;{\rm str}\,(g*f)
    =(-1)^{\pi (g)}\;{\rm str}\,(g*f) \,,
\ee
where $\pi (f)$ is the parity of $f$, i.e.
$f(-z,-y) =(-1)^{\pi (f)}f(z,y) $. This definition of parity is
in agreement with the standard spin-statistics relationship,
so that
\be
\label{str[]}
   \str \; [f,g]_* =0
\ee
for any two elements $f$ and $g$
provided that fermions carry an additional Grassmann grading.

The definition (\ref{str}) works for polynomial
elements of the star-product algebra but not necessarily makes sense
beyond this class. For example,
\be \label{stre}
    \str\; \exp{i(zy)} = \infty \,.
\ee
As a result, the supertrace (\ref{str})
is not well-defined for the algebra
of $\tilde{y}$ (\ref{tilde y}) because the integrals
in the definition of the supertrace may diverge. This is the only
point where the definition of the action is sensitive to the value of
the parameter $\nu$.
In sect.~\ref{classes} we discuss the possibility of generalization
of (\ref{str}) to the case of
arbitrary $\nu$.
The content of this section is therefore
only applicable to the particular case of $\nu = 0$. Note that, as
shown in \cite{PV}, the
equations (\ref{SS1}), (\ref{SB1}) are well-defined for any value of
$\nu$.

The Lagrangian leading to the equations (\ref{SS1}), (\ref{SB1})
has the form
\be
\label{lagrangian}
   L = i s_\ga * s^\ga * b  - 2 b  - b * b \,.
\ee
The variation of the action (\ref{action1}), (\ref{lagrangian})
w.r.t. $b$ gives rise to (\ref{SS1}) while the
variation w.r.t.  $s_\ga$ gives rise to (\ref{SB1}).
(It is the property of the supertrace (\ref{prstr})
along with the fact that $s_\ga$ carries odd parity
that leads to the anticommutator on the left hand side of (\ref{SB1})).

The field $b$ is auxiliary. It is expressed in terms of $s_\ga$
by virtue of its equation of motion (\ref{SS1}),
\be
\label{BfromS}
   b = \frac{i}{2} s_\ga * s^\ga -1  \,.
\ee
Plugging this expression back into (\ref{action1}),
(\ref{lagrangian}) we obtain an equivalent
``second-order'' action,
\be
\label{action2}
   \cS' =  {1\over 4}\; \str \left(  i s_\ga * s^\ga  - 2 \right)^2 \,.
\ee

According to the discussion in sect.~\ref{eqs}, this action
functional encodes the dynamics of massless spin-0 and spin-1/2
fields in $AdS_3$, interacting via higher spin gauge fields.
The way this dynamics is described by the action
(\ref{action1}), (\ref{lagrangian}), (\ref{action2}) is different as
compared to the standard field theoretical language since it does not
contain an explicit reference to the spacetime. Nevertheless,
according to the general ``unfolded formulation'' technics
\cite{Unf, huge} the equations (\ref{WW})-(\ref{WS})
provide a map to the usual description of the spacetime fields in
terms of appropriate partial differential equations
\cite{Eq, PV, huge}.

Remarkably, the action functional (\ref{action2}) admits
representation in the form of the two-dimensional noncommutative
Yang-Mills theory \cite{li,SW} with some infinite-dimensional
gauge group. Indeed,  let us write
\be
\label{pert}
   s_\ga = \rho z_\ga - 2i \gs_\ga
\ee
with the vacuum part $s_{0\ga} = \rho z_\ga$ and dynamical
perturbation  $-2i \gs_\ga$.
Plugging (\ref{pert}) into (\ref{action2}) and using the formula
\be
   [\rho z_\ga, f]_* = -2i \rho\; {\partial f\over \partial z^\ga}\,,
\ee
which is a simple consequence of (\ref{prod}) and (\ref{Klein}),
one obtains
\be
\label{Umod}
    i s_\ga * s^\ga = 2 - 2i\rho\,\gep^{\ga\gb}
    \left({\partial\gs_\gb\over\partial z^\ga}
   - {\partial \gs_\ga\over \partial z^\gb}\right)
   - 2i \gep^{\ga\gb}\, [\gs_\ga, \gs_\gb]_*   \,.
\ee
Introducing auxiliary spinor differentials $dz^\ga$ \cite{Eq},
\be
   dz^\ga\;dz^\gb = - dz^\gb\;dz^\ga \,,
\ee
an exterior differential
$d_z = \rho\, dz^\ga {\partial\over \partial z^\ga}$
and a spinor 1-form $\gs = dz^\ga \gs_\ga$, one
defines the field strength
\be
\label{F}
F = d_z \gs + \gs * \gs =dz^\ga dz^\gb F_{\ga\gb}.
\ee
The action
(\ref{action2}) takes a form
\be
\label{action3}
   \cS'' =- 8\; \str \left( F_{\ga\gb} * F^{\ga\gb} \right).
\ee

Let us now interpret the elements $z^\ga$ as coordinates of
a two-dimensional noncommu\-ta\-tive spacetime, identifying
$y^\ga, k, \rho, \psi_{1,2}$ with the generating elements of
the infinite-dimensional associative algebra $H$ --
the universal enveloping algebra of the commutation relations
for $y^\ga, k, \rho, \psi_{1,2}$ in (\ref{Klein}) and
(\ref{zz}). The 2-form (\ref{F}) can be identified with
the field strength in the noncommutative non-Abelian Yang-Mills
theory. The action (\ref{action3}) has a form  reminiscent of
the noncommutative non-Abelian Yang-Mills action with the
higher spin gauge algebra $H$. Here the supertrace (\ref{str})
serves both for the $z$-integration and supertrace for the gauge
algebra in the sense that it gives rise to the gauge invariant
functional of the noncommutative Yang-Mills curvatures. There
is however some difference compared to the standard situation due
to the fact that the integration measure in (\ref{str}) contains
the factor $\exp{(-i(zy))}$,
thus mixing the ``coordinate" and ``color" degrees of freedom.

Let us note that despite the fact that the Weyl algebra
with polynomial elements admits a basis
in which the supertrace factorises in the $z$ and $y$ spaces
(i.e. with the star-product algebra of polynomials in $z$ and $y$
interpreted as the tensor product of two individual algebras of
polynomials in $z$ and $y$) one
can hardly use this construction for the problem under investigation
because it admits no meaningful analog of the
nonpolynomial operator $\exp{i(zy)}$ (\ref{K})
that comes out as a result of solution of the field equations
in our construction. In other words, the same action rewritten
in the tensor product basis leads to the equations of motion that
admit no solutions.

Another important difference is that the construction we use is
essentially quantum admitting no quasiclassical (i.e. deformation)
interpretation. This is again a consequence of the relevance of
the operator $\exp{i(zy)}$ that has the form
$\exp{(i\hbar^{-1} (zy))}$ if one reintroduces
$\hbar \neq 1$ into the right hand side of the equations
(\ref{zz}). In other words, the system under investigation
admits no ``classical'' solution at $\hbar \to 0$ (see however
sect.~\ref{spec}).

Another relevant
issue is that the supertrace operation on the Weyl algebra differs
from the trace used in the deformation quantization approach. This
is a consequence of the fact
that the Weyl algebra operates with
formal power series while the deformation quantization technics
assumes integrable functions of the space-time variables that
tend to zero at infinity.

\section{Generic $\nu$}
\label{classes}

As shown in \cite{PV}, the one-parametric family of vacua
(\ref{vacua}), (\ref{tilde z}), (\ref{tilde y}) describes matter
systems with different masses parametrized by $\nu$.
The equations of motion (\ref{SS1}, \ref{SB1}) are well defined
for any value of $\nu$. However, to define the consistent action
functional for the case of general $\nu$, one should find
an appropriate generalization of the supertrace operation. In this
section we mainly set up the problem rather then solve it explicitly.
A possible interpretation of this analysis is that
the functional space describing the fluctuations
near the different vacua seem to be inequivalent.
The theories with different values of $\nu$
are somewhat reminiscent of  different
"topological" sectors, like, e.g., instantons
in the $4d$ Yang-Mills.

To see what happens, one has to analyze
the action  perturbatively. Plugging the perturbative expansions
$s_\ga= s_{0\ga} + s_{1\ga}$, $b = \nu K + b_1$
into (\ref{lagrangian}) one finds
\be
\label{S3}
L= \nu^2 + i(s_{1\ga} * s_0{}^\ga + s_{0\ga} * s_1{}^\ga) * b_1
+ i\nu\, s_{1\ga} * s_1{}^\ga * K  - b_1 * b_1 \,,
\ee
where we neglected the terms linear in $s_{1\ga}$ and
$b_1$ that do not contribute to the action
as the vacuum fields satisfy the equations of motion.
Because of the cancellation of the exponential factors contained
in the vacuum fields (\ref{vacua})-(\ref{tilde z}) with the
exponential factor in the definition of the supertrace,
the bilinear form
in the Gaussian integral with respect to the spinorial variables
in (\ref{str})
may degenerate so that the resulting expressions are not nesessarily
well-defined. In particular, after the Gaussian integration
in the spinorial integration variables in (\ref{str}) is completed,
some terms, containing the integrals in the variable $t$ arising from
(\ref{tilde z}), diverge in (\ref{S3}) at $t\to 1$. The conclusion is
that the naive extrapolation of the action to the case of arbitrary
$\nu$ is not working.

There are two main alternatives. One is that the vacuum solution
should be somehow modified  (as shown in \cite{PV}
there exists a broad class of the vacuum solutions equivalent at the
level of equations of motion). Another alternative is that one should
find a proper definition of the supertrace operation for every value
of $\nu$. We believe this latter alternative is most promising.
One argument is due to the fact that the on-mass-shell
higher spin algebras and the corresponding supertrace operations
are essentially different for different values of $\nu$.

Indeed, plugging the perturbative expansions into
the equations of motion (\ref{SS1}), (\ref{SB1})
one gets in the linear approximation
\bee
\label{SB1vac}
   s_{0\ga} * b_1 = - b_1 * s_{0\ga} \,.
\eee
As a result,
\be
\label{bkc}
b_1 = K* c\,,
\ee
where $c$ takes values in the subalgebra of elements
commuting to $s_{0\ga}$ (\ref{tilde z}), i.e.
\be
\label{A}
    c(z,y;\psi_{1,2},k)= \sum_{n=0}^\infty \frac1{n!}\;
    c_{ \ga_1\ldots\ga_n } (k, \psi_1, \psi_2)\;
    \ty^{\ga_1}*\ldots*\ty^{\ga_n}\,,
\ee
where $c_{\ga_1\ldots\ga_n}$ are totally symmetric multispinors
(i.e., we choose the Weyl ordering). This formula
is a simple consequence  \cite{PV} of (\ref{zy}) and of the
properties of the elements $k$, $\psi_1$ and $\psi_2$.

The important fact shown in \cite{Quant} is that
the universal enveloping algebra of the commutation
relations (\ref{y com}) $Aq(2,\nu)$ (in the notation of \cite{Quant})
with a generic element of the form
\be
\label{f}
    f(\ty,k) = \sum_{A=0}^1 \; \sum_{n=0}^\infty \;\frac1{n!} \;
    f^A_{\ga_1\ldots \ga_n} \;
    k^A\, \ty^{\ga_1} * \ldots * \ty^{\ga_n}  \,,
\ee
where $f^A_{\ga_1\ldots \ga_n}$ are totally symmetric multispinors,
admits a unique supertrace operation which for
 the element (\ref{f}) has the form
\be
\label{str2}
     \str_\nu\; f(\ty,k) = f^0 - \nu f^1  \,.
\ee
For $\nu=0$, it reduces to the supertrace (\ref{str}) for
$z$-independent functions.

One  way to see that this definition of supertrace
cannot be derived from the supertrace (\ref{str}) is
to observe that the following formal computation
\be
   \str\, \{k\ty_\ga \,,\ty_\gb\}_* = 2i \gep_{\ga\gb}\; \str\,(k+\nu)
   = 2i \gep_{\ga\gb}\, \nu
\ee
is in contradiction with the defining property of the
supertrace (\ref{prstr}) requiring the left hand side to be
zero. A more careful analysis shows
that this result can be interpreted as certain anomaly due
to divergences for non-polynomial functions under the
supertrace operation (\ref{str}). (Note, that, as expected,
$\str_\nu \{k\ty_\ga \,,\ty_\gb\}_* =0$.)

Of course, there is no formal contradiction with the uniqueness
theorem for the supertraces (\ref{str}) and (\ref{str2}) because
they were proved for polynomials (or formal power series).
The formulae like (\ref{tilde z}) drive us away from this class.
In \cite{PV}, it was shown that
the star-product (\ref{prod}) is well-defined for such
nonpolynomial functions. This is not true for
the supertrace (\ref{str}) however.

The question therefore is whether there exists a consistent
generalization $\str_\nu$ of the supertrace (\ref{str})
to the whole algebra, such that
it reduces to (\ref{str2}) for the functions depending only on
$\ty$ and to (\ref{str}) at $\nu =0$. One complication
is that the polynomials in $s_{0\ga}$, $\ty_\ga$,
$k$, $\psi_{1,2}$ do not form a closed algebra
as the commutation relations (\ref{z com})
contain explicitly a nonpolynomial function $K=k\, \exp{i(zy)}$
for $\nu\neq 0$.
This makes it impossible to work with the
Weyl-ordered polynomials in $s_{0\ga}$ and $\ty_\ga$ as a starting
point towards a proper generalization of (\ref{str})
\footnote{For example, one can see the difficulty from the
formal computation of
$\str_\nu\, \{kK * s_{0\ga},\; s_0{}^\ga \}_* = -4i\, \str_\nu
(\exp{i(zy)} -\nu k)$ that leads to a finite expression for
$\str_\nu\, \exp{i(zy)}$ by virtue of (\ref{str2}) that is in
contradiction with (\ref{stre}).}.
However, such an ordering prescription is not relevant for
the problem
under consideration because the star-product (\ref{prod}) is
itself defined via certain normal ordering \cite{PV,huge}.
The best we can do at this stage is to conjecture that at
$\nu \neq 0$ there exists some operation
$\str_\nu$ that reduces to (\ref{str2}), when restricted to the
subalgebra of functions of $\ty_\ga$ and $k$, and to the supertrace
operation (\ref{str}) in the limit $\nu \to 0$.
In that case, the models with different parameters $\nu$ admit the
action functionals with the same Lagrangian (\ref{lagrangian})
but with $\str_\nu$ instead of $\str$ in the action (\ref{action1}).

Let us note that if the corresponding actions exist they
may also be well-defined on-mass-shell what is
{\it a priori} neither necessary nor guaranteed. Note that the
difference between the off-mass-shell and on-mass-shell setting in
our approach is that in the former case the fluctuational parts
of the fields are arbitrary (i.e., formal) functions of the spinor
variables
$y_\ga$ and $z_\ga$, while in the latter case they depend on these
variables in a specific way governed by the field equations
(\ref{SS1}) and (\ref{SB1}) (e.g., via the dependence on
$\ty$ as in (\ref{A})). By analogy with the standard space-time
action formulations one can admit that the action functional
should not necessarily be on-mass-shell finite (for example
because of a volume divergence). Nevertheless, the insertion of
the solution (\ref{bkc}), (\ref{A}) into the action defined
with $\str_\nu$ leads to some well-defined functional by virtue
of (\ref{str2}) at least at the quadratic level. Note
that even at $\nu=0$ the variable $b$ contains via (\ref{bkc})
a factor of $\exp{i(zy)}$ on which the supertrace diverges.
Fortunately, this factor falls out of the Lagrangian
(\ref{lagrangian}) and
$\str\,L$ turns out to be well-defined at least in the quadratic
approximation in $b$. It is an interesting question even for the
case of $\nu=0$ whether this is true beyond the quadratic
approximation (i.e., at the interaction level). In fact,
the requirement that
the nonlinear action is well-defined on-mass-shell may give a
useful criterium for the selection of appropriate vacuum
solutions in connection of the locality problem discussed in
\cite{PV}.

\section{Special $\nu$}\label{spec}

It was shown in \cite{Quant} that for special values
$\nu = 2l+1$ , $l\in \bf Z$,
the algebra $Aq(2,\nu)$ acquires infinite-dimensional  ideals
$I_\nu$ identified with the null
spaces w.r.t. the invariant bilinear form constructed from the
supertrace (\ref{str2}), i.e. $\str_\nu\, (ab) = 0$,
$\forall a\in I_\nu$, $b\in Aq(2,\nu)$.
As a result, all elements that belong to $I_\nu$
do not contribute under $\str_\nu$ and therefore fall out of
the action $S = \str_\nu\, L$.
The factor algebra $Aq(2, 2l+1)/I_{2l+1}$ is isomorphic
\cite{Quant}
to the associative algebra of $(2|l|+1)\times (2|l|+1)$ matrices
treated as supermatrices from $Mat(|l|,|l|+1)$.
The supetrace $\str_\nu$ (\ref{str}) for polynomials $P(\ty )$
amounts for this case to the usual matrix supertrace.

As a result, in the $\ty_\ga$ sector (i.e., at least for the
on-mass-shell action) for the odd integer values of $\nu$
the action acquires the form of the supertrace of some
forth-order matrix polynomial. The analysis of the off-mass-shell
action requires a better understanding of the properties of
$\str_\nu$ in $s_{0\ga}$ sector. One comment is that because of
the linear dependence  of the field $s_\ga$
on $\rho$ (cf.(\ref{rho})) and nontrivial commutation relations
of $\rho$ with $s_{0\ga}$, the analysis in this sector
does not amount to the algebraic analysis of $Aq(2;\nu)$ despite
the fact that the variables $s_{0\ga}$ form themselves the
deformed oscillator algebra (\ref{z com}).
As a result there is no reason to expect
any pecularity in the $s_{0\ga}$ sector for the special values of
$\nu$. We therefore conclude that for odd integer values of
$\nu$ the action of the model becomes analogous to the
noncommutative version of Matrix String Theory action \cite{DVV}.

Let us make a few comments on the
limit $\nu\to \infty$ that leads to one more interesting
interpretation of the model. As shown in \cite{Quant},
the Lie superalgebra constructed from  $Aq(2,\nu)$
via (anti)commutators tends in this limit to the superextension
of the Lie algebra of two-dimensional volume preserving
diffeomorphisms
with the even part isomorphic to the direct sum of
two volume preserving diffeomorphism algebras of a sphere $S^2$,
$w_\infty \sim su(\infty)$. Choosing odd integer $\nu$ for
the limiting sequence $\nu\to\infty$, one recovers
the approximation of $2d$ volume-preserving
diffeomorphism algebra by matrix
algebras discovered by Hoppe \cite{Hoppe}.

The simplest way to see this correspondence is to
rewrite the commutation relations (\ref{y com})
in terms of appropriately  rescaled  variables as
\be
\label{y comin}
   [\ty_\ga,\ty_\gb]_* = 2i\gep_{\ga\gb}(\nu^{-1}+ k)\,,\quad
        \ty_\ga k=-k \ty_\ga \,.
\ee
The identification $\hbar = \nu^{-1}$ leads to the quasiclassical
interpretation of the limit $\nu \to \infty$ in terms of
two-dimensional sphere phase space. The Poisson brackets of $S^2$
coordinates
identified with the symmetric bilinear combinations of oscillators
$\ty_\ga$ do not depend on $k$ and describe $so(3)$ Lie algebra,
while the role of the $k$-dependent term in (\ref{y comin})
is that it leads to the unit radius-squared of the sphere $S^2$
given by the appropriately rescaled quadratic Casimir operator of
$so(3)$ (for more detail see \cite{Quant}).

Another important point is that, for large $\nu$,
$m\sim \nu\lambda$ where $\lambda$ is the inverse radius of $AdS_3$.
Therefore $m$ tends to infinity
in the limit $\nu\to\infty$. In other words, the sector of matter
fields decouples from the spectrum in this limit. In terms
of the coordinate-free formulation developed in this paper this
fact manifests itself as follows. The part of the field $b$
describing the matter fields is proportional
to the generating element $\psi_2$ which effectively
replaces some of the commutators in the algebra
by anticommutators due to
anticommutativity with $\psi_1$. As a result, the limit
$\nu\to\infty$ turns out to be ill-defined in this sector and
the only possibility to have $\nu=\infty$ is
to set the part of $b$ proportional to
$\psi_2$ equal to zero. This conclusion is indeed
natural in the context of analysis of \cite{PV} where it was
shown that the corresponding relativistic
matter fields acquire infinite masses in the limit $\nu\to\infty$.
The fact that the large $N$ limit in two-dimensional theory is
equivalent to the limit $m\to\infty$ in $AdS_3$ for certain matter
fields looks quite interesting.

An interesting question is whether
the limiting procedure in the $s_{0\ga}$ sector
 can be performed in such a way that the coordinates
$z^{\ga}$  would admit interpretation of commuting
coordinates of $2d$ spacetime in the limit $\nu \to \infty$.
This is not obvious because of the appearance of the terms
containing $\exp{i(zy)}$ (e.g. via $K$ in (\ref{S3})).
If nevertheless such a procedure is possible, the resulting model
may provide us with some sort of manifest realization
of the bulk/boundary correspondence \cite{maldw}. Indeed, the
original $3d$ space-time (bulk) theory containing gravity will
be equivalent
in the limit $\nu \to \infty$ to a $2d$ YM theory with an
infinite-dimensional gauge group.

\section{Conclusion}

The action proposed in this paper for $AdS_3$ higher-spin-matter
system is very different from the usual field-theoretical actions
since it is formulated in the space of auxiliary spinor variables
rather than in the usual space-time. Nevertheless, it reproduces
correctly the dynamics of the $3d$ system under consideration.
An interesting question for the future is to use
this action for quantization of the model.

For special values of the
parameter of mass the model acquires a form analogous to that
of the matrix models. It is
tempting to speculate that the relationship of our action
with the space-time higher spin actions may be analogous to the
relationship
between relativistic $M$ theory and its Matrix formulation
\cite{M}.

Note that consistent field-theoretical higher
spin actions in $AdS_4$  are known not only at the free field level
\cite{Fronsdald=4}
but also to the cubic order in interactions \cite{FV}.
The full $d \geq 4$ nonlinear higher spin gauge actions are
not yet known however.
Hopefully, the approach proposed in this paper will make it possible
to construct full nonlinear ``matrix''  actions for the
higher spin theories in $d \geq 4$.

The proposed construction exhibits a number of striking similarities
with the current ideas of the theory of fundamental interactions
associated with $M$ theory.
The $\nu \to \infty$ limit is analogous to
the $SU(\infty)$ $2d$ Yang-Mills theory
thus indicating the formal connection to Matrix String Theory
\cite{DVV}.
On the
other hand, $\nu =0$ case is proved to have a form of
the $2d$ noncommutative Yang-Mills with an infinite-dimensional
gauge group.  An interesting fact is that the large $N$ limit in
our model is equivalent to the large mass limit for
certain $AdS_3$ matter fields.

\section*{Acknowledgments}

This research was supported in part by INTAS, Grant No.96-0538,
INTAS-RFBR  Grant No.95-829
and by the RFBR Grant No.98-02-16261. S.~P. acknowledges
a partial support from Stanford University.
The research of A.~S. is supported by  the Landau
Scholarship Foundation (Forschungszentrum J\"ulich), and INTAS
Grant No. YSF 98-152.


\begin{thebibliography}{77}
\bibitem{Blen} M.~P.~Blencowe,
  {\it Class. Quantum Grav.}~{\bf 6} (1989) 443.
\bibitem{Eq} M.~A.~Vasiliev,
  {\it Mod.~Phys.~Lett.} {\bf A7} (1992) 3689.
\bibitem{PV}
  S.~F.~Prokushkin and M.~A.~Vasiliev, {\it Nucl.~Phys.} {\bf B545}
  (1999) 385, hep-th/9806236.
\bibitem{Unf} M.~A.~Vasiliev,
  {\it Class.~Quant.~Grav.} {\bf 11} (1994) 649.
\bibitem{huge} M.~A.~Vasiliev, ``Higher Spin Gauge Theories:
   Star-Product and ADS Space'', hep-th/9910096.
\bibitem{Pen} R.~Penrose,
   {\it Ann. Phys.} (N.Y.) {\bf 10} (1960) 171.
\bibitem{Penrose}
   R. Penrose and W. Rindler, {\it Spinors and space-time},
   Cambridge Univ. Press (Cambridge, 1984), Vol. 1,2.
\bibitem{Wig} E.~P.~Wigner, {\it Phys. Rep.} {\bf 77} (1950) 711.
\bibitem{Quant} M.~A.~Vasiliev,
   {\it JETP Lett.} {\bf 50} (1989) No.8, 374;
   {\it Int.~J.~Mod.~Phys.} {\bf A6} (1991) 1115.
\bibitem{BDV}
  E.~Bergshoeff, B.~de~Wit and M.~A.~Vasiliev, {\it Nucl.~Phys.} {\bf B366}
  (1991) 315.
\bibitem{Fortsch} M.~A.~Vasiliev,
   {\it Fortschr. Phys.} {\bf 36} (1988) 33.

\bibitem{li} P.-M.~Ho, Y.-S.~Wu, {\it Phys. Lett.} {\bf B398}
    (1997) 52, hep-th/9611233; {\it Phys. Rev.} {\bf D58}
    (1998) 066003, hep-th/9801147; \\
    M.~Li, {\it Nucl. Phys.} {\bf B499} (1997) 149, hep-th/9612222;
    ``Comments on Supersymmetric Yang-Mills Theory on a
    Noncommutative Torus'', hep-th/9802052; \\
    A.~Connes, M.~R.~Douglas, A.~Schwarz,
    {\it JHEP} 9802:003 (1998), hep-th/9711162; \\
    M.~R.~Douglas, C.~Hull, {\it JHEP} 9802:008 (1998),
    hep-th/9711165; \\
    L.~Cornalba, R.~Schiappa, ``Matrix Theory Star Products from
    the Born-Infeld Action'', hep-th/9907211.

\bibitem{SW} N.~Seiberg and E.~Witten, ``String Theory and
    Non-Commutative Geometry'', hep-th/9908142.
\bibitem{DVV} R. Dijkgraaf, E. Verlinde, H. Verlinde,
 {\it Nucl.Phys.} {\bf B500} (1997) 43, hep-th/9703030.

\bibitem{Hoppe} J. Hoppe, MIT PhD Thesis (1982);
Quantum theory of a relativistic surface, workshop on Constraints
theory and relativistic dynamics (Florence, 1986), eds. G. Longhi
and L. Lusanna (World Scientific, Singapore 1987) p. 267.
\bibitem{maldw} J.~Maldacena,  {\it Adv.~Theor.~Math.~Phys}
2 (1998) 231, hep-th/9711200; \\
  E.~Witten,  {\it Adv.~Theor.~Math.~Phys.} 2 (1998) 253,
  hep-th/9802150.

\bibitem{M} T. Banks, W. Fischler, S.H. Shenker, L. Susskind,
{\it Phys. Rev.} {\bf D55} (1997) 5112, hep-th/9610043; \\
T. Banks, ``TASI Lectures on Matrix Theory'', hep-th/9911068.
\bibitem{Fronsdald=4}
   C.~Fronsdal, {\it Phys. Rev.} {\bf D18} (1978) 3624; {\bf D20}
   (1979) 848;\\
   J.~Fang and C.~Fronsdal, {\it Phys. Rev.} {\bf D18} (1978) 3630;
   {\bf D22} (1980) 1361;\\
   B. de ~Wit and D.~Z.~Freedman, {\it Phys. Rev.} {\bf D21}
   (1980) 358\\
   M.~A.~Vasiliev, {\it Sov. J. Nucl. Phys.} {\bf 32} (1980) 855
   $(p.\,439$ in English translation);\\
   C.~Aragone and S.~Deser, {\it Nucl. Phys.} {\bf B170 [FS1]}
   (1980) 329;\\
   V.E. Lopatin and M.A. Vasiliev, {\it Mod. Phys. Lett.} {\bf A3}
   (1988) 257.

\bibitem{FV} E.~S.~Fradkin and M.~A.~Vasiliev, {\it Phys. Lett.}
{\bf B189} (1987) 89; {\it Nucl.~Phys.} {\bf B291} (1987) 141.


\end{thebibliography}
\end{document}